\documentclass{elsart}
\usepackage{graphicx,amsfonts,amssymb,bm,subeqnarray}
\DeclareRobustCommand\openone{\leavevmode\hbox{\small1\normalsize\kern-.33em1}}

\begin{document}

\begin{frontmatter}

\title{Effective damping in the Raman cooling of
trapped ions}

\author{A. B. Klimov {}$^{\mathrm{a}}$,
J. L. Romero {}$^{\mathrm{a}}$,
J. Delgado {}$^{\mathrm{b}}$,
L. L. S\'{a}nchez-Soto {}$^{\mathrm{b}}$}

\address{{}$^{\mathrm{a}}$
Departamento de F\'{\i}sica,
Universidad de Guadalajara,
Revoluci\'on~1500, 44420~Guadalajara,
Jalisco, Mexico}

\address{{}$^{\mathrm{b}}$
Departamento de \'Optica,
Facultad de F\'{\i}sica,
Universidad Complutense,
28040~Madrid, Spain}

\journal{Optics Communications}

\maketitle

\begin{abstract}
We present a method of treating the interaction
of a single three-level ion with two laser beams.
The idea is to apply a unitary transformation
such that the exact transformed Hamiltonian has
one of the three levels decoupled for all
values of the detunings. When one takes into
account damping, the evolution of the system
is governed by a master equation usually
obtained via adiabatic approximation under
the assumption of far-detuned lasers. To go
around the drawbacks of this technique, we use
the same unitary transformation to get an
effective master equation.
\end{abstract}

\end{frontmatter}

\section{Introduction}

In recent years, much progress has been made in
cooling and trapping techniques for neutral atoms
and for ions~\cite{Win87,Die89,Hei90,Wes90,Cas91,Kas92,Wie99}.
In fact, in a trapped ion the center-of-mass dynamics
gets entangled with the internal degrees of freedom
and some peculiar aspects of the vibrational response
have been successfully exploited for experimentally
generating Fock~\cite{Mee96}, coherent and squeezed~\cite{Lei97},
and Schr\"{o}dinger catlike states~\cite{Mon96}, proposing
theoretical schemes for engineering several nonclassical
states~\cite{Cir93,Cir94,Cir96,Mat96,Gou97,Ste97,Ger97,Mat98,Vog01,Lei03},
realizing tomographic reconstructions of the density
matrix~\cite{Wal95,Lei96,Poy96,Bar96}, and characterizing
a variety of quantum effects~\cite{Blu88,Blo92,Dut94,Hue99,Wal99}.
This is of fundamental interest, since it brings to the
forefront issues involving the relationship between
quantum and classical physics, but also offers potential
applications for, e.g., precision spectroscopy~\cite{Win97}
or quantum computation~\cite{Cir95,Chu97,Sas02,Bar03}.

In modeling typical experiments one considers a three-level
atomic system interacting with two laser fields (Raman
scheme~\cite{Vog95,Wal98}) and reduces it to a two-level
problem on the assumption of large detunings by using
the adiabatic elimination~\cite{Ors00}: the effective
Hamiltonian obtained in this way has the form of the
usual Jaynes-Cummings model.

Adiabatic elimination has been criticized on several
grounds~\cite{Pur88,Lug90,Kli99,War01}, and other methods
of deriving effective Hamiltonians exist~\cite{Kle74,Sha80}.
In this spirit, we have recently proposed an alternative approach
that involves using a unitary transformation (in fact, a nonlinear
rotation) to obtain an equivalent Hamiltonian for which one level
decouples~\cite{Kli00,Kli02}. The transformation can be exactly
found and gives the same results as the adiabatic elimination
(except for including intensity-dependent Stark shifts) when
it is evaluated up to second-order terms in coupling constants.

To take into account the effects of damping in Raman cooling schemes,
the standard way of proceeding is to start from the master
equation for the three-level system and adiabatically eliminate
the far-off resonant level. The details are described in many
different text books~\cite{Bar97,Dav76}. Unfortunately, it is
known that this treatment is not valid in many regimes of physical
interest and other approximations are
required~\cite{Sav97,Lam97,Mur98,Sch99,Rei02,Atk02}.

The main purpose of this paper is to show how our approach of
nonlinear rotations allows one to go around these drawbacks in
a natural way. Our strategy can be stated in very simple terms:
starting from the exact master equation for the three-level model,
we apply to it the same unitary transformation leading to the
effective Hamiltonian, obtaining in this way what we call an
effective master equation~\cite{Kli03}. Here we fully investigate
this approach and present numerical evidences of its validity.

\section{Physical system and model Hamiltonian}

In the interest of retaining as much clarity as possible, we
first recall some well-known facts~\cite{VWW01,Sch01} about the
system we wish to treat here, which consists of a three-level
trapped ion in the $\Lambda$ configuration with energy levels
$E_0 < E_1 < E_2$, as shown in figure 1. As usual, to describe
this system we use the operators
\begin{equation}
\hat{S}_{ij} = |j \rangle \langle i| \, ,
\end{equation}
where $|i\rangle $ denotes the eigenstate of the $i$th atomic level.
One can easily check that they satisfy
\begin{equation}
[ \hat{S}_{ij}, \hat{S}_{kl}]=
\delta _{jk} \hat{S}_{il}-
\delta_{il} \hat{S}_{kj} \, ,
\label{ccr3}
\end{equation}
which correspond to the commutation relations of the algebra
u(3). Obviously, the three ``diagonal" operators $\hat{S}_{ii}$
measure level populations, while the ``off-diagonal"
$\hat{S}_{ij}$ generate transitions from level $i$ to level $j$.

%%%%%%%%%%%%%%%%%%%%%%%%%%% FIG ! %%%%%%%%%%%%%%%%%%%%%%%%%%%%%
\begin{figure}
\centering
\resizebox{0.60\columnwidth}{!}{\includegraphics{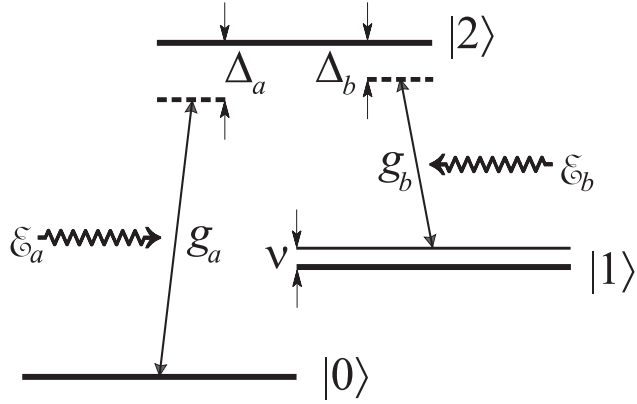}}
\caption{Energy scheme of a three-level
$\Lambda$-type ion interacting with two
laser fields, coupling the two ground states to
a common excited atomic state via a Raman transition.
Here we illustrate the case of driving the first blue
sideband.}
\end{figure}
%%%%%%%%%%%%%%%%%%%%%%%%%%%%%%%%%%%%%%%%%%%%%%%%%%%%%%%%%%%%%%%

The levels $|0 \rangle$ and $| 1 \rangle$ are
metastable and coupled by stimulated Raman transitions
via two classical optical fields (of frequencies $\omega_a$
and $\omega_b$) of the form
\begin{equation}
\mathbf{E}_\ell = \bm{\mathcal{E}}_\ell \
\exp [i (\mathbf{k}_\ell \cdot \hat{\mathbf{x}}
- \omega_\ell t - \varphi_\ell)] \, ,
\end{equation}
where from now on the index $\ell$ runs the values $a$ and $b$,
$\hat{\mathbf{x}}$ is the position operator associated with the
center-of-mass motion and $\varphi_\ell$ is the phase of the laser
field $\ell$ at the mean position of the ion.

The ion is trapped in a harmonic potential. Therefore, the
center-of-mass motion can be described in terms of annihilation
and creation operators of vibrational quanta (phonons) in the usual way
\begin{equation}
\hat{x}_q = \sqrt{\frac{\hbar}{2 M \nu_q}}
(\hat{a}_q + \hat{a}_q^\dagger ) =
\Delta x_q (\hat{a}_q + \hat{a}_q^\dagger ) \, ,
\end{equation}
where $\nu_q$ represents the oscillatory frequency along the $q$th direction,
$M$ is the ion mass, and $\Delta x_q$ is the width of the ground-state wave
function.

The Hamiltonian that describes the system is $ \hat{H} = \hat{H}_{\mathrm{cm}}
+ \hat{H}_{\mathrm{ion}} + \hat{V}$, where
\begin{eqnarray}
\hat{H}_{\mathrm{cm}} & = &
\sum_q  \hbar \nu_q \
\hat{a}_q^\dagger \hat{a}_q \, , \nonumber \\
\hat{H}_{\mathrm{ion}} & = &
\sum_i E_i \ \hat{S}_{ii}  \, ,  \\
\hat{V} & = &  \hbar [ g_a(\hat{\mathbf{x}}) \ e^{- i \omega_a t} \hat{S}_{02} +
g_a^\ast (\hat{\mathbf{x}}) \ e^{i \omega_a t} \hat{S}_{20}]
\nonumber \\
& + & \hbar
[ g_b(\hat{\mathbf{x}}) \ e^{- i \omega_b t} \hat{S}_{12} +
g_b^\ast (\hat{\mathbf{x}}) \ e^{i \omega_b t} \hat{S}_{21} ] \, .
\nonumber
\end{eqnarray}
The interaction term $\hat{V}$ is written in the rotating-wave
approximation and the coupling constants are
\begin{equation}
g_\ell (\hat{\mathbf{x}}) = \kappa_\ell
\mathcal{E}_\ell \ \exp[ i ( \mathbf{k}_\ell
\cdot \hat{\mathbf{x}} - \varphi_\ell ) ] \, ,
\end{equation}
where $\kappa_\ell$ is the corresponding dipole matrix element in
the direction of the driving field.

Because one has the trivial constraint $\hat{S}_{00} + \hat{S}_{11} +
\hat{S}_{22}= \hat{\openone}$, only two populations can vary independently.
Eliminating the population of the level $| 2 \rangle $ we can recast
the Hamiltonian as $ \hat{H} = \hat{H}_0 + \hat{H}_{\mathrm{int}}$, with
\begin{eqnarray}
\label{H0}
\hat{H}_0 & = & \sum_q  \hbar \nu_q
\hat{a}_q^\dagger \hat{a}_q  -
\hbar (\Delta_a + \omega_a) \hat{S}_{00} -
\hbar (\Delta_b + \omega_b) \hat{S}_{11}  \, ,
\nonumber \\
& &  \\
\hat{H}_{\mathrm{int}} & = & \hbar
[ g_a(\hat{\mathbf{x}}) \ e^{- i \omega_a t} \hat{S}_{02} +
g_a^\ast (\hat{\mathbf{x}}) \ e^{i \omega_a t} \hat{S}_{20}]
\nonumber \\
& + &
\hbar
[ g_b(\hat{\mathbf{x}}) \ e^{- i \omega_b t} \hat{S}_{12} +
g_b^\ast (\hat{\mathbf{x}}) \ e^{i \omega_b t} \hat{S}_{21} ] \, ,
\nonumber
\end{eqnarray}
where we have defined the following detunings
\begin{equation}
\hbar \Delta_a  =  E_2 - E_0 - \hbar \omega_a \, ,
\qquad
\hbar \Delta_b  =  E_2 - E_1 - \hbar \omega_b \, .
\end{equation}
This Hamiltonian contains terms oscillating rapidly in time at
frequencies $\omega_a$ and $\omega_b$, which can be eliminated
by going to a rotating frame. The final result is
\begin{eqnarray}
\label{H0til}
\hat{H}_0 & = & \sum_q  \hbar \nu_q \
\hat{a}_q^\dagger \hat{a}_q  -
\hbar (\Delta_a \hat{S}_{00} +
\Delta_b  \hat{S}_{11} )  \, , \nonumber \\
& &  \\
\hat{H}_{\mathrm{int}} & = & \hbar
[ g_a(\hat{\mathbf{x}})  \hat{S}_{02} +
g_a^\ast (\hat{\mathbf{x}})  \hat{S}_{20}]  +
\hbar [ g_b(\hat{\mathbf{x}})  \hat{S}_{12} +
g_b^\ast (\hat{\mathbf{x}}) \hat{S}_{21} ] \, .
\nonumber
\end{eqnarray}
This is the basic Hamiltonian that will be used in our subsequent
analysis.

\section{Effective Raman Hamiltonian in the dispersive limit}

The standard treatment assumes that level $| 2 \rangle$ is far
off resonance and proceeds via adiabatic elimination to obtain
an effective two-level Raman interaction Hamiltonian with an
intensity-dependent coupling between levels $|0 \rangle$ and
$|1 \rangle$. In Refs.~\cite{Kli00} and \cite{Kli02} we have
claimed that such a procedure has drawbacks and proposed instead
an alternative technique involving nonlinear rotations. For
the problem at hand, we introduce the unitary transformation
\begin{equation}
\label{H1}
\hat{\mathcal{H}} =
\hat{T} \hat{H} \hat{T}^\dagger \, ,
\end{equation}
where $\hat{T} = \exp[\hat{J}(\hat{\mathbf{x}})]$ and
\begin{equation}
\label{rotat}
\hat{J} (\hat{\mathbf{x}})  = [
\varepsilon_a (\hat{\mathbf{x}}) \hat{S}_{02} -
\varepsilon_a^\ast (\hat{\mathbf{x}}) \hat{S}_{20} ]
+ [ \varepsilon_b (\hat{\mathbf{x}}) \hat{S}_{12} -
\varepsilon_b^\ast (\hat{\mathbf{x}}) \hat{S}_{21} ] \, .
\end{equation}
Here the parameters $\varepsilon_a (\hat{\mathbf{x}})$
and $\varepsilon_b (\hat{\mathbf{x}})$ are defined by
\begin{equation}
\varepsilon_\ell (\hat{\mathbf{x}}) =
\frac{g_\ell (\hat{\mathbf{x}})}{\Delta_\ell} \, .
\end{equation}
To interpret this operator $\hat{T}$, we note that
($\hat{S}_{02}, \hat{S}_{20}$) and ($\hat{S}_{12},
\hat{S}_{21}$) are raising and lowering su(2) operators
that correspond to the allowed transitions
$0 \leftrightarrow 2$ and $1 \leftrightarrow 2$. However,
these two dipoles are not independent, since Eq.~(\ref{ccr3})
imposes highly nontrivial couplings between them. In consequence,
$\hat{T}$ can be seen as a ``rotation" acting on the subspace
of these two  dipoles.

By using the well-known expression
\begin{equation}
\label{sum}
e^{\hat{A}} \hat{B} e^{-\hat{A}} = \sum_{n=0}^\infty
\frac{\hat{B}^{(n)}}{n!} \, ,
\end{equation}
where $\hat{B}^{(n)} = [\hat{A}, \hat{B}^{(n-1)} ]$, and
$\hat{B}^{(0)} = \hat{B}$, the exact transformation law (\ref{H1})
has been found in Refs.~\cite{Wu96} and \cite{Wu97} (see
also \cite{Ale95}). The explicit expression is complicated
although is valid for any values of the detunings.  Since we
are assuming that level $| 2 \rangle$ is far off resonance and,
consequently, the ratios $\varepsilon_a (\hat{\mathbf{x}})$ and
$\varepsilon_b (\hat{\mathbf{x}})$ can be taken as small
quantities, the series (\ref{sum}) can be evaluated keeping only
up to second-order terms. By applying this to (\ref{H0til}) we
finally obtain an effective Hamiltonian $\hat{\mathcal{H}} =
\hat{\mathcal{H}}_0 + \hat{\mathcal{H}}_{\mathrm{int}}$, where
\begin{eqnarray}
\hat{\mathcal{H}}_0 & = & \sum_{q}
\hbar \nu_q \ \hat{a}_q^\dagger  \hat{a}_q -
\hbar \left( \Delta_a + \frac{|g_a (\hat{\mathbf{x}}) |^2}
{\Delta_a} \right ) \hat{S}_{00} - \hbar \left( \Delta_b +
\frac{|g_b (\hat{\mathbf{x}} ) |^2}
{\Delta_b} \right ) \hat{S}_{11} \, , \nonumber \\
& & \\
\hat{\mathcal{H}}_{\mathrm{int}} & = &
- \frac{\hbar}{2} \left ( \frac{1}{\Delta_a} +
\frac{1}{\Delta_b} \right) [ g_a (\hat{\mathbf{x}})
g_b^\ast (\hat{\mathbf{x}})  \hat{S}_{01}+
g_a^\ast (\hat{\mathbf{x}}) g_b (\hat{\mathbf{x}})
\hat{S}_{10} ] \, . \nonumber
\end{eqnarray}

In the usual experiments the wave vector difference is
chosen to be parallel with the $X$ direction of the trap,
so $(\mathbf{k}_a - \mathbf{k}_b) \cdot \hat{\mathbf{x}} =
\delta k \ \hat{x}$ and the interaction couples only the
motion in $X$ direction to the internal state of the trapped
ion. Using the effective inversion between levels $| 1 \rangle$
and $| 0 \rangle$:
\begin{equation}
S_{01}^z  = \frac{1}{2} ( \hat{S}_{11}- \hat{S}_{00} ) \, ,
\end{equation}
we can write
\begin{subeqnarray}
\label{Heff0}
\hat{\mathcal{H}}_0 & = &  \hbar \nu  \
\hat{a}^\dagger \hat{a} + \hbar \delta \ \hat{S}_{01}^z \, ,  \\
& & \nonumber \\
\label{Heffint}
\hat{\mathcal{H}}_{\mathrm{int}} & = & -
\frac{\hbar \Omega}{2} ( e^{i  \delta k \, \hat{x}} \
\hat{S}_{01} + e^{- i  \delta k \, \hat{x}} \ \hat{S}_{10} ) \, ,
\end{subeqnarray}
with
\begin{equation}
\delta  =  \Delta_a -\Delta_b +
\frac{| g_a |^2}{\Delta_a}-
\frac{| g_b |^2}{\Delta_b} \, ,
\qquad
\Omega  =  |g_a g_b| \left ( \frac{1}{\Delta_a} +
\frac{1}{\Delta_b} \right) \, .
\end{equation}
In terms of the phonon raising and lowering operators we
rewrite (\ref{Heffint}b) as
\begin{equation}
\label{temp}
\hat{\mathcal{H}}_{\mathrm{int}} =
\frac{\hbar \Omega}{2} [ e^{i \eta
(\hat{a} + \hat{a}^\dagger) } \ \hat{S}_{01} +
e^{- i \eta (\hat{a} + \hat{a}^\dagger) } \
\hat{S}_{10} ] \, ,
\end{equation}
where the Lamb-Dicke parameter is defined as
\begin{equation}
\label{LDick}
\eta = \delta k \sqrt{\frac{\hbar}{2 M \nu}} \, ,
\end{equation}
and represents the ratio between the recoil energy and
the quantum vibrational energy, both taken in the $X$ direction.
It is worth observing that the second-order corrections
to this effective Hamiltonian vanish, so (\ref{temp})
accurately describes the system dynamics up to times
$t \leqslant \hbar / (g_\ell \varepsilon_\ell^3)$.

In the interaction picture relative to $\hat{\mathcal{H}}_0$
we finally get
\begin{equation}
\label{Haa}
\hat{\mathcal{H}}_{\mathrm{int}}  =
- \frac{\hbar \Omega}{2}
\{  e^{ i [ \eta (\hat{a} + \hat{a}^\dagger) -\delta t] } \
\hat{S}_{01} +
e^{- i [ \eta ( \hat{a} + \hat{a}^\dagger) - \delta t] } \
\hat{S}_{10} \} \, .
\end{equation}
By tuning the frequency difference $\delta$ to an integer
multiple of the trap frequency $\nu$, $\delta = (n^\prime - n ) \nu$,
we can resonantly drive transitions from $|0, n \rangle$ to
$|1, n^\prime \rangle$, where the ket $|j, n \rangle$ indicates
the $n$th vibrational Fock state in the electronic state $j$
($j = 0, 1$). In this case, $\hat{\mathcal{H}}_{\mathrm{int}}$
is dominated by a single stationary term. The exponent
$\exp [ i\eta ( \hat{a} + \hat{a}^\dagger ) ]$ in Eq.~(\ref{Haa})
contains all powers of $\hat{a}$ and $\hat{a}^\dagger$. However,
all contributions with $m  \neq n^\prime - n$ oscillate rapidly
and average out when $\nu$ is much larger than $\Omega$. We assume
the Lamb-Dicke limit, in which $\eta \sqrt{\bar{n} + 1} \ll 1$.
In the relevant case of the first red sideband $\delta = \nu$,
we get to lowest order in $\eta$
\begin{equation}
\label{JCH}
\hat{\mathcal{H}}_{\mathrm{int}}   =
- i \eta \frac{\hbar \Omega}{2}
( \hat{a} \hat{S}_{01} -  \hat{a}^\dagger \hat{S}_{10} ) \, ,
\end{equation}
which is the familiar Jaynes-Cummings Hamiltonian.
Similarly, there is a first blue sideband,
corresponding to an anti-Jaynes-Cummings Hamiltonian
and higher-order sidebands, but in the rest of
this paper we shall be mainly concerned with the model
Hamiltonian (\ref{JCH}).

\section{Damping in terms of an effective master equation}

For many purposes, the coherent control of the vibrational
dynamics plays a crucial role. For the model discussed in the
previous section damping effects have been observed that even
occurred under almost ideal conditions~\cite{Win98}.

The presence of the two lasers causes the appearance of a
coupling between internal (electronic) and external (center of
mass) degrees of freedom of the trapped ion. However, when the
Lamb-Dicke parameter is very small, the dynamics due to this
coupling is slow compared with the internal dynamics that may
be adiabatically eliminated. This reduction leads to a master
equation for the motional degrees of freedom, where the involved
transition rates depend on steady-state expectation values of
internal operators~\cite{Ors00}.

To take full advantage of the method outlined in the previous
section, we start from the density operator $\hat{\rho}$ for
the external and internal degrees of freedom of the three-level
ion and make the hypothesis that the dynamics is described by
a general master equation of the Lindblad type~\cite{Lin76}:
\begin{equation}
\label{meqor}
\frac{d}{dt} \hat{\rho} =
\frac{1}{i \hbar} [ \hat{H}, \hat{\rho} ] +
\frac{\gamma_a}{2} \mathcal{L}[\hat{S}_{20}] \ \hat{\rho}
+ \frac{\gamma_b}{2} \mathcal{L}[\hat{S}_{21}] \ \hat{\rho} \, ,
\end{equation}
where $\gamma_a$ and $\gamma_b$ represent the decoherence
rates for the processes associated with the coupling of the
dipoles with a zero-temperature bath and $\mathcal{L}[\hat{C}]$
is the Lindblad superoperator
\begin{equation}
\mathcal{L}[\hat{C}] \ \hat{\rho} =
2 \hat{C} \hat{\rho} \hat{C}^\dagger -
\{ \hat{C}^\dagger \hat{C}, \hat{\rho}\} \, .
\end{equation}
Note that (\ref{meqor}) describes an irreversible evolution of
the system at different rates for each channel. It is implicitly
assumed that the both dipole moments are orthogonal to each other;
that is, $\mathbf{d}_{02} \cdot \mathbf{d}_{12}^\ast = 0$, where
$\mathbf{d}_{ij}$ are the transition dipole matrix elements~\cite{Aga74}.

It seems natural to ask how this equation is transformed by the same
unitary operator $\hat{T}$ leading to the effective Hamiltonian in (\ref{H1}).
Let us denote the effective density matrix by
\begin{equation}
\hat{\varrho} = \hat{T}  \hat{\rho}  \hat{T}^\dagger \, .
\end{equation}
Taking into account that $ \hat{S}_{20}\hat{\rho} =
\hat{S}_{21}\hat{\rho} =0$, we get up to second-order terms
\begin{eqnarray}
\label{2oter}
\hat{T} \hat{S}_{02} \hat{T}^\dagger   & = &
\hat{S}_{02} - \varepsilon_a^\ast (\hat{x}) \hat{S}_{00} -
\varepsilon_b^\ast(\hat{x}) \hat{S}_{01} \, , \nonumber \\
& & \\
\hat{T} \hat{S}_{12} \hat{T}^\dagger  & = &
\hat{S}_{12} -\varepsilon_b^\ast (\hat{x}) \hat{S}_{11}-
\varepsilon_a^\ast (\hat{x}) \hat{S}_{10} \, . \nonumber
\end{eqnarray}
Then, if we apply $\hat{T}$ to (\ref{meqor}), we obtain
the effective master equation
\begin{eqnarray}
\frac{d}{dt} \hat{\varrho} & = &
\frac{1}{i \hbar} [ \hat{\mathcal{H}}_{\mathrm{int}},
\hat{\varrho} ] \nonumber \\
& + &  \frac{\gamma_a}{2} \mathcal{L} [ \varepsilon_a (\hat{x}) \hat{S}_{00}
+ \varepsilon_b (\hat{x}) \hat{S}_{10} ] \ \hat{\varrho} \nonumber \\
& + & \frac{\gamma_b}{2} \mathcal{L} [ \varepsilon_b (\hat{x}) \hat{S}_{11}
+ \varepsilon_a (\hat{x}) \hat{S}_{01} ] \ \hat{\varrho} \, ,
\end{eqnarray}
where $\hat{\varrho}$ is expressed in the interaction picture.

In the Lamb-Dicke limit we can make the approximation
\begin{equation}
\varepsilon_\ell (\mathbf{x} ) \simeq \varepsilon_\ell ( 1 +
\eta_\ell \hat{x} ) \, ,
\end{equation}
with [compare with (\ref{LDick})]
\begin{equation}
\eta_\ell = k_\ell \sqrt{\frac{\hbar}{2 M \nu}} \, ,
\end{equation}
and the master equation takes the simpler form
\begin{eqnarray}
\label{meqf}
\frac{d}{dt} \hat{\varrho} & = &
[\hat{\mathcal{H}}_{\mathrm{int}}, \hat{\varrho} ]
\nonumber \\
& + & \frac{\gamma_a |\varepsilon_a |^2}{2}
\mathcal{L} [ \hat{S}_{00} ] \ \hat{\varrho} +
\frac{\gamma_b |\varepsilon_b|^2}{2}
\mathcal{L} [ \hat{S}_{11} ] \ \hat{\varrho}
\nonumber \\
& + & \frac{\gamma_a |\varepsilon_b|^2}{2}
\mathcal{L} [ \hat{S}_{10} ] \ \hat{\varrho}
+ \frac{\gamma_b |\varepsilon_a|^2}{2}
\mathcal{L} [ \hat{S}_{01} ] \ \hat{\varrho} \nonumber \\
& + & \frac{\gamma_a}{2}
\mathcal{K} [ \varepsilon_a (\hat{x}) \hat{S}_{00},
\varepsilon_b (\hat{x}) \hat{S}_{10} ] \ \hat{\varrho}
+ \frac{\gamma_b}{2} \mathcal{K} [ \varepsilon_b (\hat{x}) \hat{S}_{11},
\varepsilon_a (\hat{x}) \hat{S}_{01} ] \ \hat{\varrho} \, .
\end{eqnarray}
This equation has a very suggestive and transparent
physical meaning: the terms $\mathcal{L} [ \hat{S}_{00}] \
\hat{\varrho}$ and $\mathcal{L} [ \hat{S}_{11} ] \
\hat{\varrho}$ describe pure phase dissipation, meanwhile
the terms $ \mathcal{L} [ \hat{S}_{10}] \ \hat{\varrho}$
and $\mathcal{L} [ \hat{S}_{01} ] \ \hat{\varrho}$ describe
dissipative-like transitions from the level $| 0 \rangle$ to
$| 1 \rangle$ and back, which leads to a stationary distribution
of population in these levels (incoherent mixing). These two terms
seem to simulate the coupling to a thermal bath at finite temperature,
but there is a strong difference: now the corresponding rates
for the processes from level $| 0 \rangle$ to $| 1 \rangle$ and
from level $| 1 \rangle$ to $| 0 \rangle$ are not related by a
Maxwell-Boltzmann factor of the form $n / (n +1)$. All these terms
have a purely atomic nature and appear in a similar master equation
derived in a different context by Di Fidio and Vogel~\cite{Fid00},
who interpreted them in terms of quantum jumps.

Besides, we have also ``crossed terms" described by superoperators
$\mathcal{K}$, which contribute substantially (i.e., the corresponding
terms are time independent in the rotating frame) only when some resonance
conditions discussed after Eq.~(\ref{Haa}) hold. For the first red sideband
we are considering here, they take the form
\begin{eqnarray}
\label{crosst}
\mathcal{K} [ \varepsilon_a ( \hat{x} ) \hat{S}_{00},
\varepsilon_b ( \hat{x} ) \hat{S}_{10} ] \ \hat {\varrho}  & = &  i
\varepsilon_a \varepsilon_b^\ast \left [ 2 ( \eta_a \hat{S}_{00}
\hat{a} \hat{\varrho} \hat{S}_{01}- \eta_b \hat{S}_{00}
\hat{\varrho} \hat{a} \hat{S}_{01} ) \right . \nonumber  \\
& - & \left . \eta ( \hat{a} \hat{S}_{01} \hat{S}_{00} \hat{\varrho} +
\hat{\varrho} \hat{a} \hat{S}_{01} \hat{S}_{00}  ) + \mathrm{h.\ c.}
\right ] \, , \nonumber \\
&& \\
\mathcal{K} [\varepsilon_b  ( \hat{x} ) \hat{S}_{11},
\varepsilon_a ( \hat{x} ) \hat{S}_{01} ] \ \hat {\varrho} & = & i
\varepsilon_b \varepsilon_a^\ast \left [ 2 ( \eta_b \hat{S}_{11}
\hat{a}^\dagger \hat{\varrho} \hat{S}_{10} - \eta_a \hat{S}_{11}
\hat{\varrho} \hat{a}^\dagger \hat{S}_{10} ) \right . \nonumber  \\
& + & \left . \eta ( \hat{a}^\dagger \hat{S}_{10} \hat{S}_{11}
\hat{\varrho} + \hat{\varrho} \hat{a}^{\dagger} \hat{S}_{10}
\hat{S}_{11} ) + \mathrm{h.\ c.} \right ] \, , \nonumber
\end{eqnarray}
where we have retained only the dominant terms in the
parameter $\eta= \eta_a - \eta_b$.

To test our theory, we have numerically integrated the master
equation (\ref{meqf}) using the Quantum Optics Toolbox produced
by S. M. Tan~\cite{Tan}. In the typical experiments at NIST~\cite{Lei96},
a single ${}^9$Be$^+$ ion is stored in a RF Paul trap with a secular frequency
along $X$ of $\nu/2\pi \simeq$ 11.2~MHz, providing a spread of the ground state
wave function of  $\Delta x \simeq $ 7~nm, with a Lamb-Dicke parameter of
$\eta \simeq 0.202$. The two laser beams, with 0.5 W in each one, are approximately
detuned $\Delta/2\pi \simeq$ 12 GHz, so that  $\Omega/2\pi \simeq$ 475~kHz.
With these data we find $\varepsilon_a \sim \varepsilon_b \simeq 0.01$, so they
can be considered as small parameters, as assumed in the previous Section.
We take also $(\gamma_a + \gamma_b) /2 \pi \simeq$ 19.4~MHz (which is about
one linewidth of the transition) and $\gamma_a =  \gamma_b$. The observable
measured in all these experiments is the fluorescence signal, which
is the probability $P_\downarrow (t)$ of occupation of the electronic
level $|0 \rangle$. This probability may be written as
\begin{equation}
P_\downarrow (t) = \sum_n
\langle 0, n | \hat{\varrho}(t) | 0, n \rangle \, .
\end{equation}

%%%%%%%%%%%%%%%%%%%%%%%%%%% FIG 3 %%%%%%%%%%%%%%%%%%%%%%%%%%%%%
\begin{figure}
\centering
\resizebox{0.75\columnwidth}{!}{\includegraphics{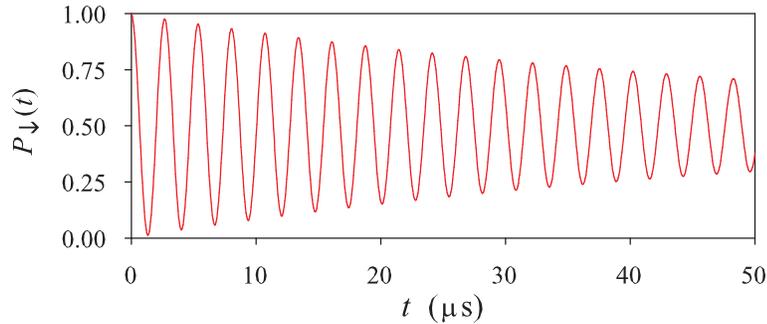}}
\caption{Evolution of the $P_\downarrow (t)$ for an initial
Fock state with $n_0=1$ driven by a Raman interaction.
The parameters are the same as in the experiment~\cite{Mee96}:
$\nu/2\pi \simeq$ 11.2~MHz, $\Delta/2\pi \simeq$ 12 GHz,
$\eta \simeq 0.202$, $\Omega/2\pi \simeq$ 475~kHz, and
$(\gamma_a + \gamma_b) /2 \pi \simeq$ 19.4~MHz with
$\gamma_a = \gamma_b$.}
\end{figure}
%%%%%%%%%%%%%%%%%%%%%%%%%%%%%%%%%%%%%%%%%%%%%%%%%%%%%%%%%%%%%%%

Let us consider first the case in which the ion starts in a
Fock state with $n_0$ excitations. The result for $n_0 =1$
appears in Fig.~2, showing clearly the existence of
damped Rabi oscillations. We have numerically checked that
in this case the role played by the ``crossed terms"
(\ref{crosst}) is insignificant. Similar results have
been found in Ref.~\cite{Fid00} by solving the
master equation with quantum trajectory methods~\cite{Car93,Cas96,Ple98}.
The advantage of this approach is that the damping
can be understood without the need of introducing
phenomenological noise. Stochastic models~\cite{Jam98,Sch98},
leading to a dispersivelike decoherence dynamics~\cite{Bud02},
have also been used for the same reason. We stress that
our theory gives essentially the same results by resorting
only to two pure Lindblad terms of very easy interpretation.

An intriguing result found in the experiments of Ref.~\cite{Mee96}
is that the fluorescence signal, for initial Fock states,
may be approximately modeled by
\begin{equation}
P_\downarrow (t) \simeq
\frac{1}{2} \left [ 1 + \cos (2 \Omega_{n_0} t)
e^{ - \gamma_{n_0} t} \right ] \, ,
\end{equation}
where $\Omega_{n_0}$ is the associated Rabi frequency
and the phenomenological decay constants $\gamma_{n_0}$
were fitted as $\gamma_{n_0} \simeq \gamma_0 (n_0 + 1)^{0.7}$.
This exponential decay can be inferred with good accuracy
from a numerical analysis of our simulated data~\cite{Bud02}.

%%%%%%%%%%%%%%%%%%%%%%%%%%% FIG 3 %%%%%%%%%%%%%%%%%%%%%%%%%%%%%
\begin{figure}
\centering
\resizebox{0.75\columnwidth}{!}{\includegraphics{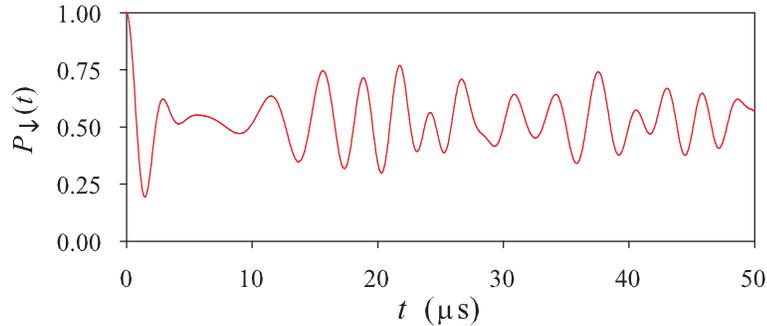}}
\caption{Evolution of the $P_\downarrow (t)$ for an initial
coherent state with an average vibrational number $\bar{n} = 3$
driven by a Raman interaction. The rest of the parameters are
the same as in Fig.~2.}
\end{figure}
%%%%%%%%%%%%%%%%%%%%%%%%%%%%%%%%%%%%%%%%%%%%%%%%%%%%%%%%%%%%%%%

In Fig.~3 we show our results for the evolution of the
fluorescence signal for an initial coherent state with an
average vibrational number $\bar{n} = 3$. The graphic
reproduces all the salient features of the experiment ~\cite{Lei96},
although for a perfect fitting a more precise value of our
parameters $\eta_a$ and $\eta_b$ would be needed.
The influence of the terms~(\ref{crosst}) is again very small.

%%%%%%%%%%%%%%%%%%%%%%%%%%% FIG 3 %%%%%%%%%%%%%%%%%%%%%%%%%%%%%
\begin{figure}[b]
\centering
\resizebox{0.75\columnwidth}{!}{\includegraphics{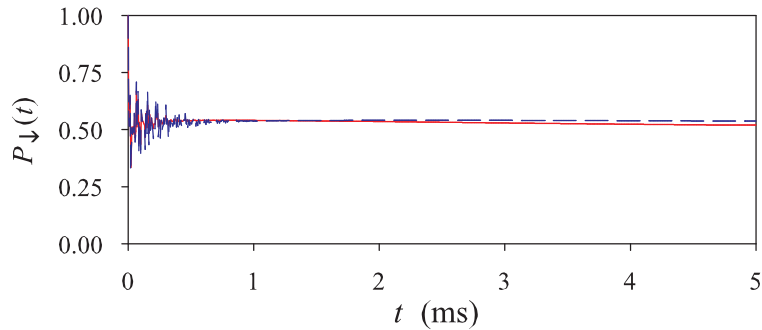}}
\caption{Evolution of the $P_\downarrow (t)$ for the same initial
coherent state with an average vibrational number $\bar{n} = 3$
as in Fig. 3 in a different time scale. The solid line represents
the case $\gamma_a/\gamma_b=1$, while the dashed line is for
 $\gamma_a/\gamma_b=3$.}
\end{figure}
%%%%%%%%%%%%%%%%%%%%%%%%%%%%%%%%%%%%%%%%%%%%%%%%%%%%%%%%%%%%%%%

In Fig.~4 we have plotted  the dynamics of the
same coherent state with $\bar{n}= 3$, but in a larger
time scale and for two different values of $\gamma_a/\gamma_b$.
We see that the stationary limit of the oscillations is not
0.5, due to the phenomenon of incoherent mixing mentioned above:
while the details of the collapse are almost insensitive
to the values of the ratio $\gamma_a / \gamma_b$, as time goes
by the quasi-stationary values of the population tend to be
different. This cannot be reproduced by using the standard
master equation approach~\cite{Lei03}. The ``crossed terms"
are important here in preventing that for larger values of
$\gamma_a / \gamma_b$ the solutions would differ too much.

%%%%%%%%%%%%%%%%%%%%%%%%%%% FIG 3 %%%%%%%%%%%%%%%%%%%%%%%%%%%%%
\begin{figure}
\centering
\resizebox{0.75\columnwidth}{!}{\includegraphics{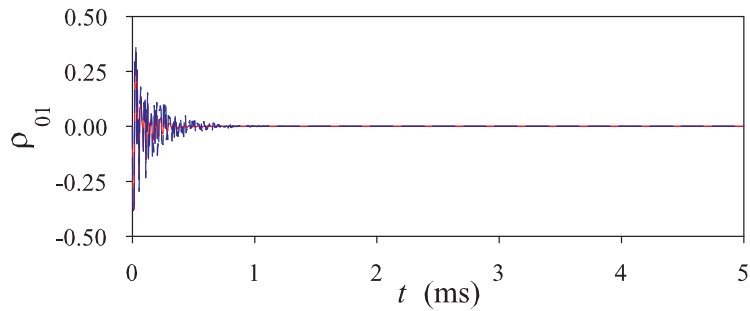}}
\caption{Evolution of the coherences $\langle \hat{\varrho}_{01} \rangle$
for the state as in Fig.~4.}
\end{figure}
%%%%%%%%%%%%%%%%%%%%%%%%%%%%%%%%%%%%%%%%%%%%%%%%%%%%%%%%%%%%%%%

To confirm that this is, in fact, an incoherent effect that affects
only to the population dynamics, in Fig.~5 we show the evolution of
the coherence $\hat{\varrho}_{01}$ for the same initial state, and
we clearly see that the values of $\gamma_a / \gamma_b$ do not influence
at all to the dynamics at the large of the dipole moment.

\section{Concluding remarks}

What we expect to have accomplished in this paper is to present
a comprehensive method of treating two-photon stimulated Raman
transitions in a single trapped ion. Our approach is based on
the application of a unitary transformation and leads in a natural
way to an effective master equation of the Lindblad type with a
clear physical interpretation.

In the framework of this description, we have shown that the evolution
of populations and dipole moments predicted by our theory is in good
agreement with the data of realistic experiments. The appearance of an
incoherent mixing that leads to a stationary redistribution of populations
has been also studied.

\section*{Acknowledgements}

We are very grateful to one anonymous reviewer for interesting comments
and suggestions on the manuscript.


\begin{thebibliography}{99}

\bibitem{Win87}
D. J. Wineland, W. M. Itano,
J. C. Bergquist and R. G. Hulet,
Phys. Rev. A \textbf{36}, 2220 (1987).

\bibitem{Die89}
F. Diedrich, J. C. Bergquist,
W. M. Itano and D. J. Wineland,
Phys. Rev. Lett. \textbf{62}, 403 (1989).

\bibitem{Hei90}
D. J. Heinzen and D. J. Wineland,
Phys. Rev. A \textbf{42}, 2977 (1990).

\bibitem{Wes90}
C. I. Westbrook, R. N. Watts, C. E. Tanner,
S. L. Rolston, W. D. Phillips, P. D. Lett
and P. L. Gould,
Phys. Rev. Lett. 65, 33 (1990).

\bibitem{Cas91}
Y. Castin and J. Dalibard,
Europhys. Lett. \textbf{14}, 761 (1991).

\bibitem{Kas92}
M. Kasevich and S. Chu,
Phys. Rev. Lett. \textbf{69}, 1741 (1992).

\bibitem{Wie99}
C. E. Wieman, D. E. Pritchard and D. J. Wineland,
Rev. Mod. Phys. \textbf{71}, S253 (1999).

\bibitem{Mee96}
D. M. Meekhof, C. Monroe, B. E. King,
W. M. Itano and D. J. Wineland,
Phys. Rev. Lett. \textbf{76}, 1796 (1996).

\bibitem{Lei97}
D. Leibfried, D. M. Meekhof, C. Monroe,
B. E. King, W. M. Itano and D. J. Wineland,
J. Mod. Opt. \textbf{44}, 2485 (1997).

\bibitem{Mon96}
C. Monroe, D. M. Meekhof, B. E. King
and D. J. Wineland,
Science \textbf{272}, 1131 (1996).

\bibitem{Cir93}
J. I. Cirac, R. Blatt, A. S. Parkins and P. Zoller,
Phys. Rev. Lett. \textbf{70}, 762 (1993).

\bibitem{Cir94}
J. I. Cirac, R. Blatt, A. S. Parkins and P. Zoller,
Phys. Rev. A \textbf{49}, 1202 (1994).

\bibitem{Cir96}
J. I. Cirac, A. S. Parkins,
R. Blatt and P. Zoller,
Phys. Rev. Lett. \textbf{70}, 556 (1996).

\bibitem{Mat96}
R. L. de Matos Filho and W. Vogel,
Phys. Rev. Lett. \textbf{76}, 608 (1996).

\bibitem{Gou97}
S. C. Gou, J. Steinbach and P. L. Knight,
Phys. Rev. A \textbf{55}, 3719 (1997).

\bibitem{Ste97}
J. Steinbach, J. Twamley and P. L. Knight,
Phys. Rev. A \textbf{56}, 4815 (1997).

\bibitem{Ger97}
C. C. Gerry, S. C. Gou and J. Steinbach,
Phys. Rev. A \textbf{55}, 630 (1997).

\bibitem{Mat98}
R. L. de Matos Filho and W. Vogel,
Phys. Rev. A \textbf{58}, R1661 (1998).

\bibitem{Vog01}
W. Vogel and S. Wallentowitz,
\textit{Manipulation of the quantum state of a trapped ion},
in Coherence and Statistics of Photons and Atoms, ed. by J. Perina
(Wiley, New York, 2001).

\bibitem{Lei03}
D. Leibfried, R. Blatt, C. Monroe and D. Wineland,
Rev. Mod. Phys. \textbf{75}, 281 (2003).

\bibitem{Wal95}
S. Wallentowitz and W. Vogel,
Phys. Rev. Lett. \textbf{75}, 2932 (1995).

\bibitem{Lei96}
D. Leibfried, D. M. Meekhof, B. E. King, C. Monroe,
W. M. Itano and D. J. Wineland,
Phys. Rev. Lett. \textbf{77}, 4281 (1996).

\bibitem{Poy96}
J. F. Poyatos, R. Walser, J. I. Cirac,
P. Zoller and R. Blatt,
Phys. Rev. A \textbf{53}, R1966 (1996).

\bibitem{Bar96}
P. J. Bardoff, C. Leichtle, G. Schrade and
W. P. Schleich,
Phys. Rev. Lett. \textbf{77}, 2198 (1996).

\bibitem{Blu88}
R. Bl\"{u}mel, J. M. Chen, E. Peik, W. Quint,
W. Schleich, Y. R. Shen and H. Walther,
Nature \textbf{334}, 309 (1988).

\bibitem{Blo92}
C. A. Blockley, D. F. Walls and H. Risken,
Europhys. Lett. \textbf{77}, 509 (1992).

\bibitem{Dut94}
S. M. Dutra, P. L. Knight and H. Moya-Cessa,
Phys. Rev. A \textbf{49}, 1993 (1994).

\bibitem{Hue99}
R. Huesmann, Ch. Balzer, Ph. Courteille,
W. Neuhauser and P. E. Toschek,
Phys. Rev. Lett. \textbf{82}, 1611 (1999).

\bibitem{Wal99}
S. Wallentowitz, W. Vogel and P. L. Knight,
Phys. Rev. A \textbf{59}, 531 (1999).

\bibitem{Win97}
D. J. Wineland, J. J. Bollinger W. M. Itano,
F. L. Moore and D. J. Heinzen,
Phys. Rev. A \textbf{46}, R6797 (1997).

\bibitem{Cir95}
J. I. Cirac and P. Zoller,
Phys. Rev. Lett. \textbf{74}, 4091 (1995).

\bibitem{Chu97}
I. L. Chung and Y. Yamamoto,
Phys. Rev. A \textbf{55}, 115 (1997).

\bibitem{Sas02}
M. \v{S}a\v{s}ura and V. Bu\v{z}ek,
J. Mod. Opt.   \textbf{49} 1593 (2002).

\bibitem{Bar03}
M. D. Barrett, B. DeMarco, T. Schaetz,
D. Leibfried, J. Britton, J. Chiaverini,
W. M. Itano, B. Jelenkovi\v{c}, J. D. Jost,
C. Langer, T. Rosenband and D. J. Wineland,
Phys. Rev. A \textbf{68}, 042302 (2003).


\bibitem{Vog95}
W. Vogel and R. L. de Matos Filho,
Phys. Rev. A \textbf{52}, 4214 (1995).

\bibitem{Wal98}
S. Wallentowitz and W. Vogel,
Phys. Rev. A \textbf{58}, 679 (1998).


\bibitem{Ors00}
M. Orszag, \textit{Quantum Optics}
(Springer, New York, 2000).

\bibitem{Pur88}
R. R. Puri and R. K. Bullough,
J. Opt. Soc. Am. B \textbf{5}, 2021 (1988).

\bibitem{Lug90}
L. A. Lugiato, P.  Galatola and  L. M. Narducci,
Opt. Commun. \textbf{76}, 276 (1990).

\bibitem{Kli99}
A. B. Klimov, J. Negro, R. Farias and  S. M. Chumakov,
J. Opt. B \textbf{1}, 562 (1999).

\bibitem{War01}
P. Warszawski and H. M. Wiseman,
Phys. Rev. A \textbf{63}, 013803 (2001).

\bibitem{Kle74}
D. J. Klein,
J. Chem. Phys. \textbf{61}, 786 (1974).

\bibitem{Sha80}
I. Shavitt and  L. T. Redmon,
J. Chem. Phys. \textbf{73}, 5711 (1980).

\bibitem{Kli00}
A. B. Klimov and  L. L. S\'anchez-Soto
Phys. Rev. A \textbf{61}, 063802 (2000).

\bibitem{Kli02}
A. B. Klimov, L. L. S\'anchez-Soto,
A. Navarro and E. C. Yustas,
J. Mod. Opt. \textbf{49}, 2211 (2002).

\bibitem{Bar97}
S. M. Barnett and P. M. Radmore,
\textit{Methods in Theoretical Quantum Optics}
(Oxford University Press, Oxford, 1997).

\bibitem{Dav76}
E. B. Davies,
\textit{Theory of Open Systems}
(Academic, London, 1976).
\bibitem{Sav97}
T. A. Savard, K. M. O'Hara and J. E. Thomas,
Phys. Rev. A \textbf{56}, R1095 (1997).

\bibitem{Lam97}
S. K. Lamoreaux,
Phys. Rev. A \textbf{56}, 4970 (1997).

\bibitem{Mur98}
M. Murao and P. L. Knight,
Phys. Rev. A \textbf{58}, 663 (1998).

\bibitem{Sch99}
S. Schneider and G. J. Milburn,
Phys. Rev. A \textbf{59}, 3766 (1999).

\bibitem{Rei02}
D. Rei\ss, K. Abich, W. Neuhauser,
Ch. Wunderlich and P. E. Toschek,
Phys. Rev. A \textbf{65}, 053401 (2002).

\bibitem{Atk02}
D. J. Atkins, H. M. Wiseman and
P. Warszawski,
Phys. Rev. A \textbf{67}, 023802 (2003).

\bibitem{Kli03}
A. B. Klimov, J. L. Romero, J. Delgado and
L. L. S\'anchez-Soto,
J. Opt. B \textbf{5}, 34 (2003).

\bibitem{VWW01}
W. Vogel, D.-G. Welsch and S. Wallentowitz,
\textit{Quantum Optics. An Introduction}
(Wiley-VCH, Berlin, 2001, 2nd edition).

\bibitem{Sch01}
W. P. Schleich,
\textit{Quantum Optics in Phase Space}
(Wiley-VCH, Weinheim, 2001).


\bibitem{Wu96}
Y. Wu,
Phys. Rev. A \textbf{54}, 1586 (1996).

\bibitem{Wu97}
Y. Wu and X. Yang,
Phys. Rev. A \textbf{56}, 2443 (1997).

\bibitem{Ale95}
M. Alexanian and S. K. Bose,
Phys. Rev. A \textbf{52}, 2218 (1995).


\bibitem{Win98}
D. J. Wineland, C. Monroe, W. M. Itano,
D. Leibfried, B. E. King and D. M. Meekhof,
J. Res. Natl Inst. Stand. Technol.
\textbf{103}, 259 (1998).

\bibitem{Lin76}
G. Lindblad,
Commun. Math. Phys. \textbf{48}, 119 (1976).

\bibitem{Aga74}
G. S.Agarwal,
\textit{Quantum Statistical Theories of
Spontaneous Emission and their Relation to
Other Approaches} (Springer-Verlag, Berlin, 1974).

\bibitem{Fid00}
C. Di Fidio and W. Vogel,
Phys. Rev. A \textbf{62}, 031802(R) (2000).

\bibitem{Tan}
S. M. Tan,
\textit{Quantum Optics Toolbox and
accompanying User Guide}, available at
the URL
http://www.phy.auckland.ac.nz/Staff/smt/qotoolbox/download.html.

\bibitem{Car93}
H. J. Carmichael,
\textit{An Open Systems Approach to Quantum Optics}
(Springer-Verlag, Berlin, 1993).

\bibitem{Cas96}
Y. Castin and K. Molmer,
Phys. Rev. A \textbf{54}, 5275 (1996).

\bibitem{Ple98}
M. B. Plenio and P. L. Knight,
Rev. Mod. Phys. \textbf{70}, 101 (1998).

\bibitem{Jam98}
D. F. James,
Phys. Rev. Lett. \textbf{81}, 317 (1998).

\bibitem{Sch98}
S. Schneider and G. J. Milburn,
Phys. Rev. A \textbf{57}, 3766 (1999).

\bibitem{Bud02}
A. A. Budini, R. L. de Matos Filho and N. Zagury,
Phys. Rev. A \textbf{65}, 041402(R) (2002).
\end{thebibliography}
\end{document}